\documentclass[onecolumn,preprintnumbers,elsart]{revtex4}

\usepackage{amsmath}
\usepackage{cases}
\usepackage{mathrsfs}
\usepackage{graphicx}
\usepackage{dcolumn}
\usepackage{bm}
\usepackage[center]{subfigure}
\usepackage{color}

\begin{document}

\title{Control of the stability and soliton formation of dipole moments in a nonlinear plasmonic finite nanoparticle array}

\author{Zhijie Mai$^{1,2}$}
\author{Shenhe Fu$^{1}$}
\author{Yongyao Li$^{2}$}
\author{Xing Zhu$^{1}$}
\author{Yikun Liu$^{1}$}
\author{Juntao Li$^{1}$}
\email{lijt3@mail.sysu.edu.cn}

\affiliation{$^{1}$State Key Laboratory of Optoelectronic Materials and Technologies,\\
School of Physics and Engineering, Sun Yat-sen University, Guangzhou 510275, China\\
$^{2}$Department of Applied Physics, South China Agricultural University, Guangzhou 510642, China\\
}

\begin{abstract}
We perform numerical analysis of a finite nanoparticle array, in which the transversal dipolar polarizations are excited by a homogenous optical field. Considering the linearly long-range dipole-dipole interaction and the cubic dipole nonlinearity of particle, the characteristics of stability of a finite number nanoparticle array should be revised, compared with that of an infinite number nanoparticle array. A critical point in the low branch of the bistable curve is found, beyond which the low branch becomes unstable for a finite number of nanoparticles. The influence of the external field intensities and detuning frequencies on this critical point are investigated in detail. When the total number of particles approaches infinity, our results become similar to that of an infinity number particle system \cite{oe}. Notably, with appropriate external optical field, a dark dipole soliton is formed. Moreover, when the scaled detuning is set to an appropriate value, a double monopole dark soliton (DMDS) consisting of two particles is formed. The DMDS may have potential applications in the subwavelength highly precise detection because of its very small width.\\
\end{abstract}

\maketitle
\section{Introduction}

Localized plasmon resonance \cite{Maier} is the excitation of the conduction electrons of metallic nanostructures coupled to an electromagnetic field. This effect leads to field amplification both inside and outside (in the near-field zone) the metal nanoparticles. These resonances, which may be used in optical sensing, generally depend on the type of metal, the shape of nanoparticles and the dielectric environment within the electromagnetic near field [2-7]. Nanoparticles are also used to constitute a subwavelength waveguide. Metallic  nanoparticles arranged into arrays can spatially confine and manipulate optical energy in a distance much smaller than the wavelength[8-12]. In addition, the strong geometric confinement of the optical field in these nanoparticle arrays can boost the efficiency of nonlinear optical effects such as the frequency conversion, modulation of optical signals, and the formation of solitons\cite{martti}.

Nonlinear plasmonics \cite{martti} which involves the excitation of nanostructures by an external field, is an emerging research field. In practice, using the field penetration inside the nanostructures to generate the resonance plasmonic, many optical nonlinear effects can be boosted\cite{sr15}. These effects include self-phase modulation in structured nanoparticles arrays \cite{sr16}, second and third harmonic generation in nanostructured metal films and nanoantennas [16-22], subwavelength solitons in different nanostructures including multilayers[23-25], nanowires arrays [26-28], Kerr nonlinear coupled-cavity array \cite{egorov}, and two-dimensional lattices \cite{noskov2013}. Recently, considering the linear dipole-dipole interaction, cubic dipole nonlinearity, and other related environment parameters, Noskov et al. analyzed the modulation instability and bistability of optical-induced dipoles and discussed several novel nonlinear effects, which include domain walls as well as the bright and dark oscillons and solitons in a metallic spherical nanoparticle array [31-33]. They performed the stability analysis using an infinite number particle system. We continue the previous works, considering the finite system, and achieve some interesting results, two of which are: first, demonstrating in detail the difference of instability areas between the infinite and finite systems consisted of nanoparticle arrays. From a practical point of view, the system of finite nanoparticle arrays should be considered and the edge effects must be addressed, both of which have been considered in this paper; second, illustrating the width variation of dipole solitons with different initial conditions. Particularly, with suitable settings, we achieve dipole solitons with extremely narrow width.

This paper is organized as follow. The stability analysis for this system is discussed in section II. Based on the analysis in section II, we propose a scheme to generate dark dipole solitons and control the width of such solitons in section III. We demonstrate that the width of a dark soliton can be controlled by an external field and different detuning frequencies. Moreover, when the scaled detuning is set to $-0.11\leq\Omega\leq-0.08$, two-particle double monopole dark soliton (DMDS), with an approximate width of 30 nm, is achieved. The paper is concluded in section IV.

\section{Model and stability analysis}
We consider the system consisted of identical spherical silver nanoparticles which are arrayed linearly equidistant with each other and embedded into an SiO$_{2}$ host (seen from Fig. 1). Specific parameters of the system are shown in the appendix.

\begin{figure}[htbp]
\centering\includegraphics[width=10cm,height=5cm]{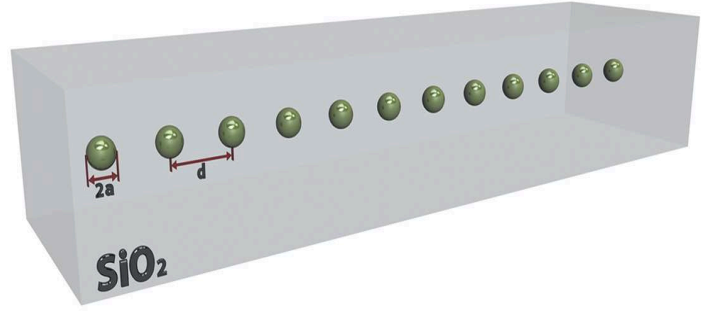}\label{fig1}
\caption{(Color online) Schematic sketch of an array of silver spherical nanoparticles embedded into an SiO$_{2}$ host, radius of sphere is $a$ and the center-to-center distance is $d$.}
\end{figure}

The nonlinear dynamic of dipoles induced by the field in the nanoparticle arrays can be described by the following equations [31-33]:
\begin{eqnarray}
-i\frac{dP_{n}^{\perp}}{d\tau}+(-i\gamma+\Omega+|P_{n}|^{2})P_{n}^{\perp}+\sum_{m\neq n}G_{n,m}^{\perp}P_{m}^{\perp}&=&E_{n}^{\perp}, \label{diff1} \\
-i\frac{dP_{n}^{\parallel}}{d\tau}+(-i\gamma+\Omega+|P_{n}|^{2})P_{n}^{\parallel}+\sum_{m\neq n}G_{n,m}^{\parallel}P_{m}^{\parallel}&=&E_{n}^{\parallel}. \label{diff2}
\end{eqnarray}
In Eqs. (\ref{diff1})and (\ref{diff2}), $P_{n}^{\bot, \parallel}=p_{n}^{\perp,\parallel}\sqrt{\chi^{(3)}}/(\sqrt{2(\varepsilon_{\infty}+2\varepsilon_{h})}\varepsilon_{h}a^3)$ are the dimensionless slow-varying amplitudes of the vertical and parallel dipoles of the $n^{\mathrm{th}}$ particle, respectively. The indices `$\perp$' and `$\parallel$' represent the vertical and parallel directions with respect to the array axis, respectively. Thus the total intensity of the dipole for the $n^{\mathrm{th}}$ particle is given as $|P_{n}|^2=|P_{n}^{\perp}|^2+|P_{n}^{\parallel}|^2$. $\gamma=\nu/(2\omega_{0})+(k_{0}a)^3\varepsilon_{h}/(\varepsilon_{\infty}+2\varepsilon_{h})$, with $k_{0}=\omega_{0}/c\sqrt{\varepsilon_{h}}$ is the scaled damping. $\Omega=(\omega-\omega_{0})/\omega_{0}$ is the detuning frequency of the dipoles. When incident wavelength is about $400nm$, $\Omega=0$ for silver particle. $\tau=\omega_{0}t$ is the scaled elapse time. $E_{n}^{\perp, \parallel}=-3\varepsilon_{h}\sqrt{\chi^{(3)}}E_{n}^{ex, \perp, \parallel}/\sqrt{8(\varepsilon_{\infty}+2\varepsilon_{h})^3}$ are also the slow varying amplitudes of the external optical fields in the respective directions. $G_{n,m}^{\perp,\parallel}$ is the linearly coupled parameter between the $n^{\mathrm{th}}$ and $m^{\mathrm{th}}$ particles in the corresponding directions and is induced by the long-range dipole-dipole interactions. $G_{n,m}^{\perp,\parallel}$ can be expressed:
\begin{eqnarray}
G_{n,m}^{\perp}&=&\frac{\eta}{2}\left[(k_{0}d)^{2}-\frac{ik_{0}d}{|n-m|}-\frac{1}{|n-m|^2}\right]\frac{e^{-ik_{0}d|n-m|}}{|n-m|},\\
G_{n,m}^{\parallel}&=&\eta\left(\frac{ik_{0}d}{|n-m|}+\frac{1}{|n-m|^2}\right)\frac{e^{-ik_{0}d|n-m|}}{|n-m|},
\end{eqnarray}
where $\eta=\frac{3\varepsilon_{h}}{\varepsilon_{\infty}+2\varepsilon_{h}}(\frac{a}{d})^3$.

Eqs. (\ref{diff1}) and (\ref{diff2}) are suitable for the case of finite or infinite nanoparticle chain. The real-time evolution methods can be used to solve the $P_{n}$ in Eqs. (\ref{diff1}) and (\ref{diff2}).\\

From Ref. \cite{oe}, we only consider the optically induced dipole in the vertical direction (i.e., the direction of `$\perp$'). Therefore, the dipole of a particle can be written as $\mathbf{P}_{n}=(P^{\perp}_{n},P^{\parallel}_{n})=(P_{n},0)$. The stationary solution of Eq. (\ref{diff1}) with infinite number of particles can be given by the following stationary equation, which is assumed to be stimulated by the vertical polarized homogenous optical field $\mathbf{E}_{n}=(E_{0}^{\perp},E_{0}^{\parallel})=(E_{0},0)$
\begin{equation}
(-i\gamma+\Omega+\sum_{j=1}^{\infty}A_{j}^{\perp}+|P_{0}|^{2})P_{0}=E_{0}, \label{st_eq}
\end{equation}
where $P_{0}$ is the stationary solution and
\begin{equation}
A_{j}^{\perp}=\eta(-1/j^3-ik_{0}d/j^2+(k_{0}d)^2/j)\exp(-ik_{0}dj),
\end{equation}
where $A_{j}^{\perp}$ is the transition from $G_{n,m}^{\perp}$ by replacing $|n-m|=j$, which considers the symmetrical structure of the nanoparticles \cite{oe}. The stationary solution in Eq. (\ref{st_eq}) can be solved numerically by dividing the stationary dipole $P_{0}$ and electric field into real and imaginary parts.
\begin{eqnarray}
P_{0}&=&P_{1}+iP_{2}, \\
E_{0}&=&E_{1}+iE_{2}. \label{RI_PE}
\end{eqnarray}
Substituting Eq. (\ref{RI_PE}) into Eq. (\ref{st_eq}), two nonlinear equations can be obtained with respect to $P_{1}$ and $P_{2}$. We adopt $E_{1}=E_{2}$, such treatment is convenient to introduce the phase of optical field in numerical simulation. But we found that the phase of $E_{0}$ does not affect the final results.  Figs 2(a) and 2(d) display typical examples of the intensities of the solutions in Eq. (\ref{st_eq}) as functions of $|E_{0}|^{2}$ with different values of $\Omega$. The function curve of $\left|P_{0}(|E_{0}|)\right|^{2}$ is drawn as a standard bistable curve, which is constructed by three branches, namely, the low, middle, and top branches. For an infinite number of particles, according to the description in Ref. \cite{oe}, the stability of the solution can be identified through the growth rate of the small perturbation noises, and the analytical results show that the solutions are stable at the low and top branches, but unstable at in the middle branch.

However, for a finite number of particle system, being considered in our work, is of significance for the experimental realization. In this system, a new instability source may be created by the finite volume effect. Consequently, the stability of the solutions should be verified again under this case. In the stability analysis based on the growth rate of the small perturbations illustrated in Ref. \cite{oe} is built from the infinite model. Thus, we use the direct simulation [real-time evolution of Eq. (\ref{diff1})] in the finite model to test the stability of the solutions. Free boundary condition can be used here and $n$ is set to a value. Our results show that the solutions of the top and the middle branches of the bistable curve are the same as those of an infinite system. By contrast, for the low branch, the front part of the curve is stable, and the latter part of the curve, which links to the middle branch, becomes unstable [Figs. 2(a) and 2(d)]. Hence, a critical point of $|E_{0}|^{2}$ separates these two parts in the low branch. The instability area of the finite system is obviously expanded by the edge effects. As an example depicted in Fig.2(a), the instability area in the low branch occupies nearly 40\% of the total low branch, indicating the strong influence of edge effects to the instability area.

Figs. 2(b,e) and (c,f) display the result of the direct simulations to the solutions in the vicinities of the left and right sides of the critical point with $\Omega=-0.1$ and $-0.2$ after the evolution of $\tau=2000$, respectively. Typically, in Figs. 2 (b) and (e), the results show that these solutions are stable, while in Figs. 2(c) and (f), the value of the dipole moments eventually jump onto the top branches after the evolution, which shows that the solutions under this circumstance are obviously unstable.

\begin{figure}[htbp]
\centering
\subfigure[]{\includegraphics[width=5cm, height=4cm]{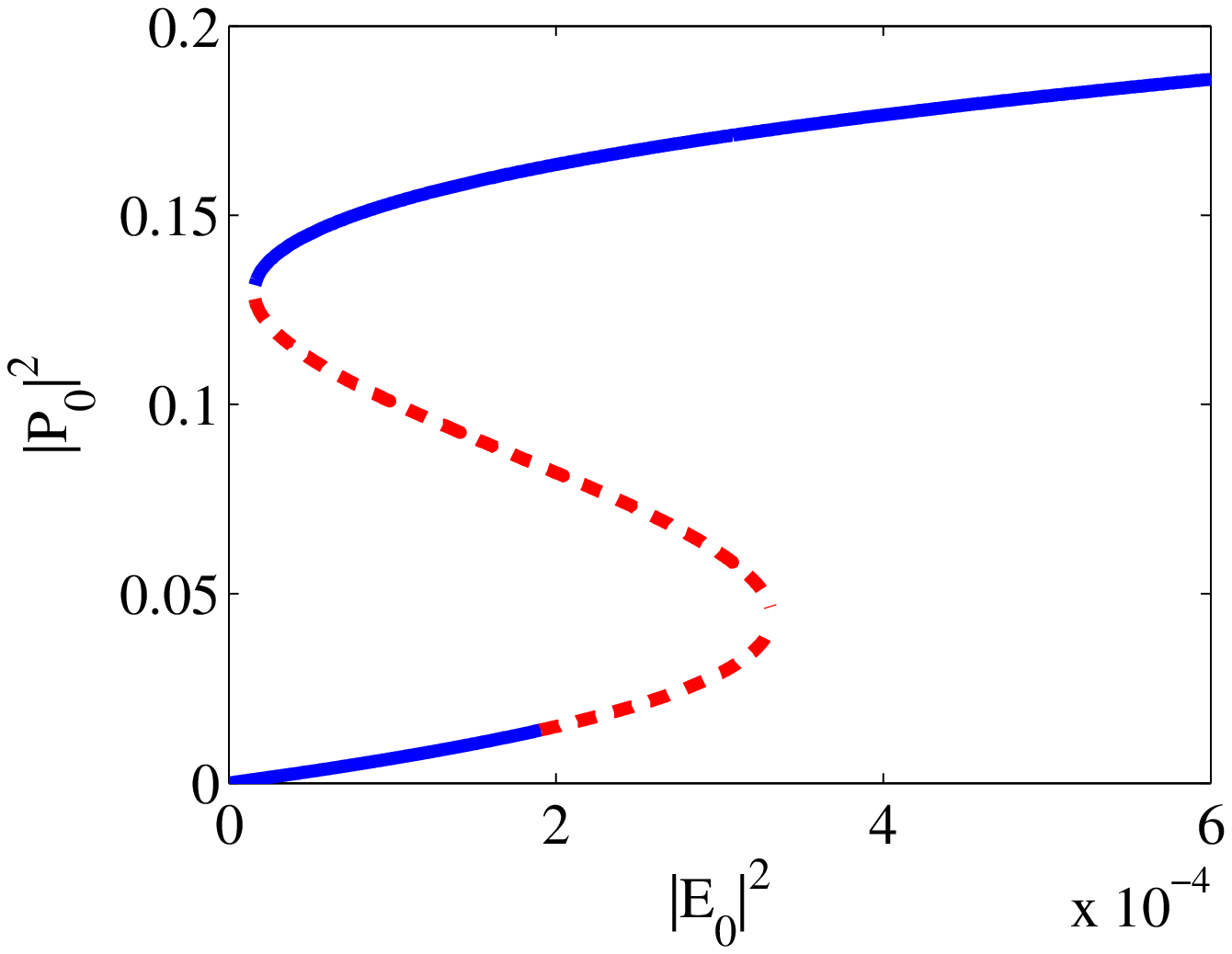}}\label{fig2a}
\subfigure[]{\includegraphics[width=5cm, height=4cm]{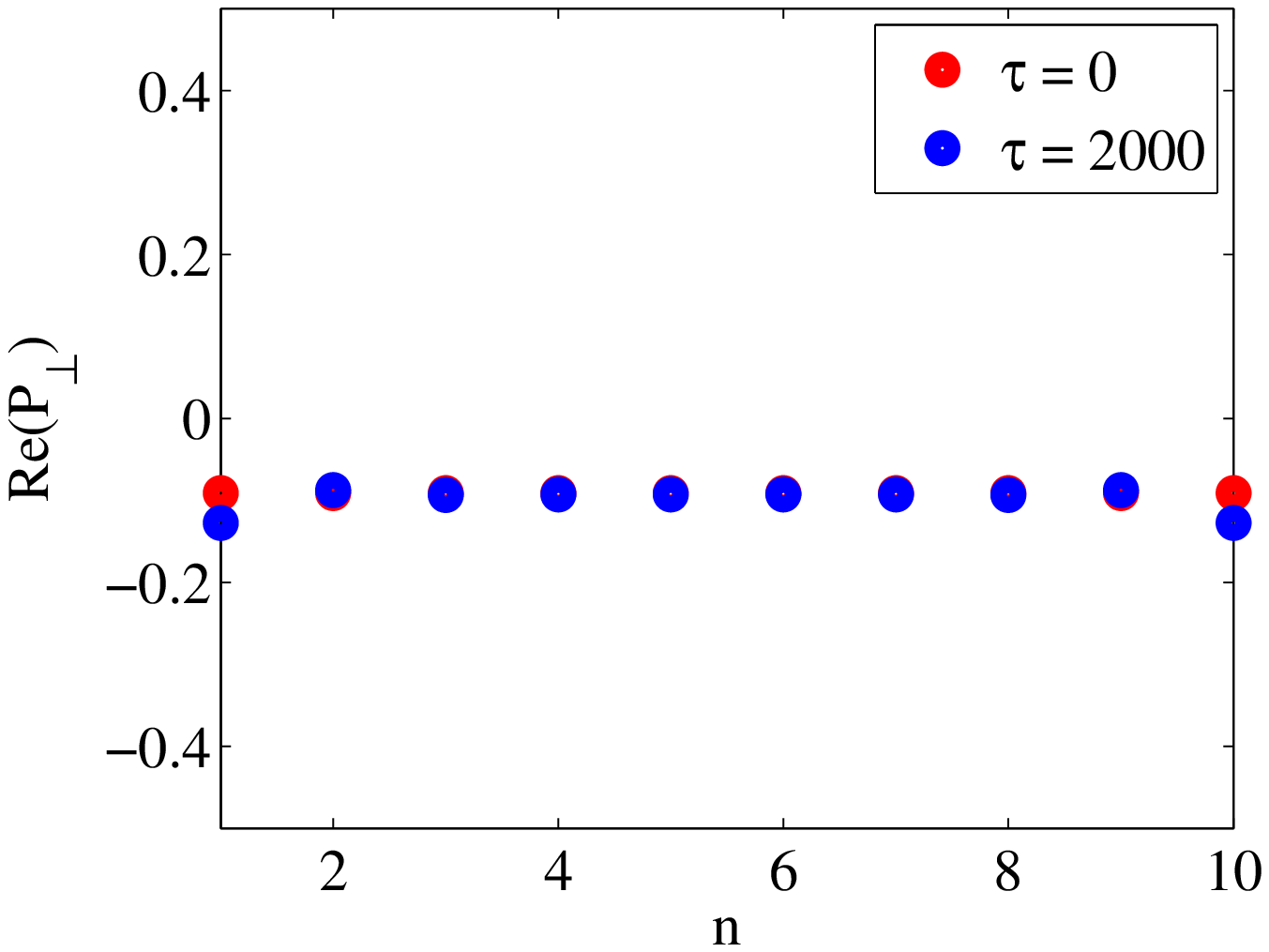}}\label{fig2b}
\subfigure[]{\includegraphics[width=5cm, height=4cm]{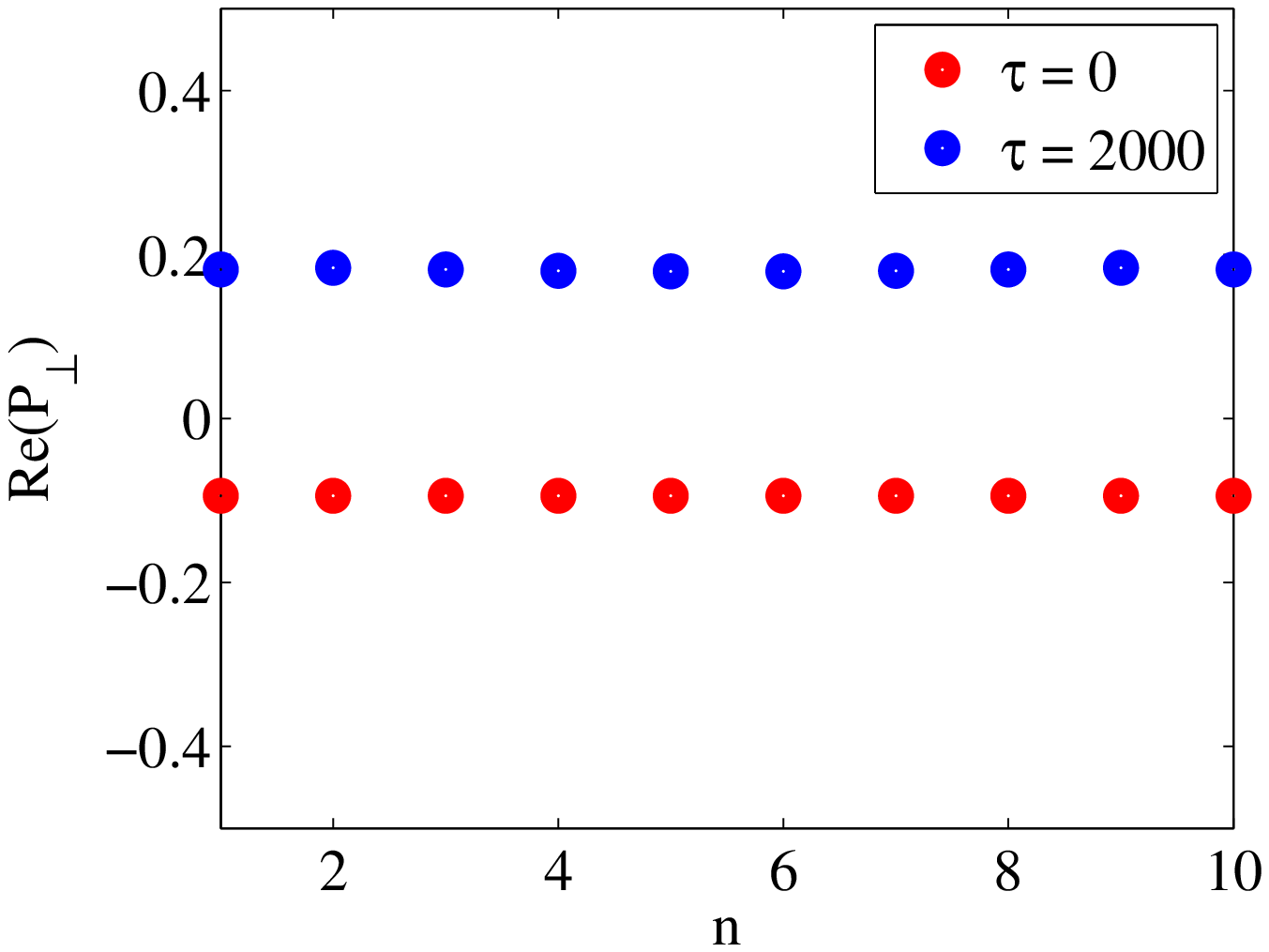}}\label{fig2c}\\
\subfigure[]{\includegraphics[width=5cm, height=4cm]{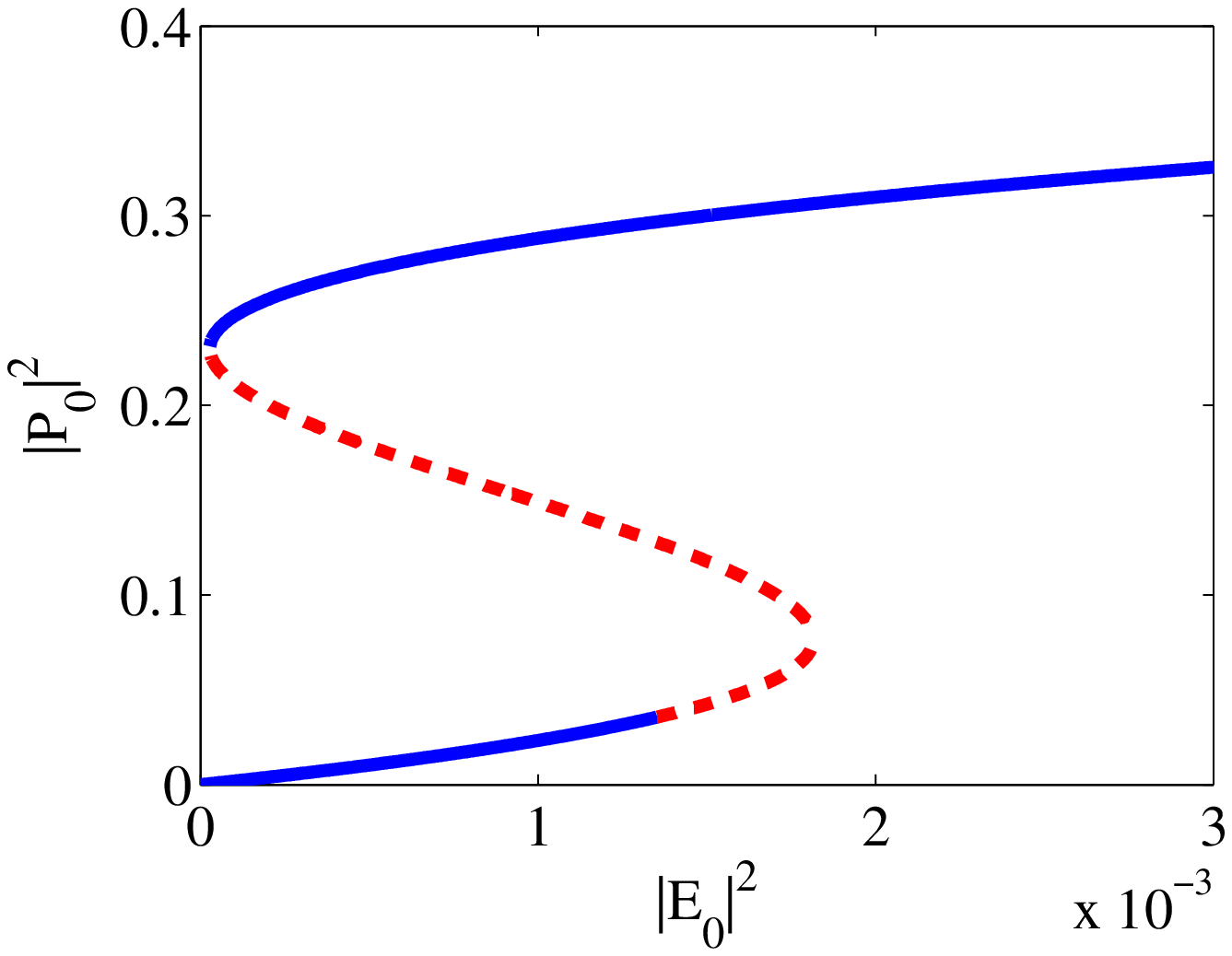}}\label{fig2d}
\subfigure[]{\includegraphics[width=5cm, height=4cm]{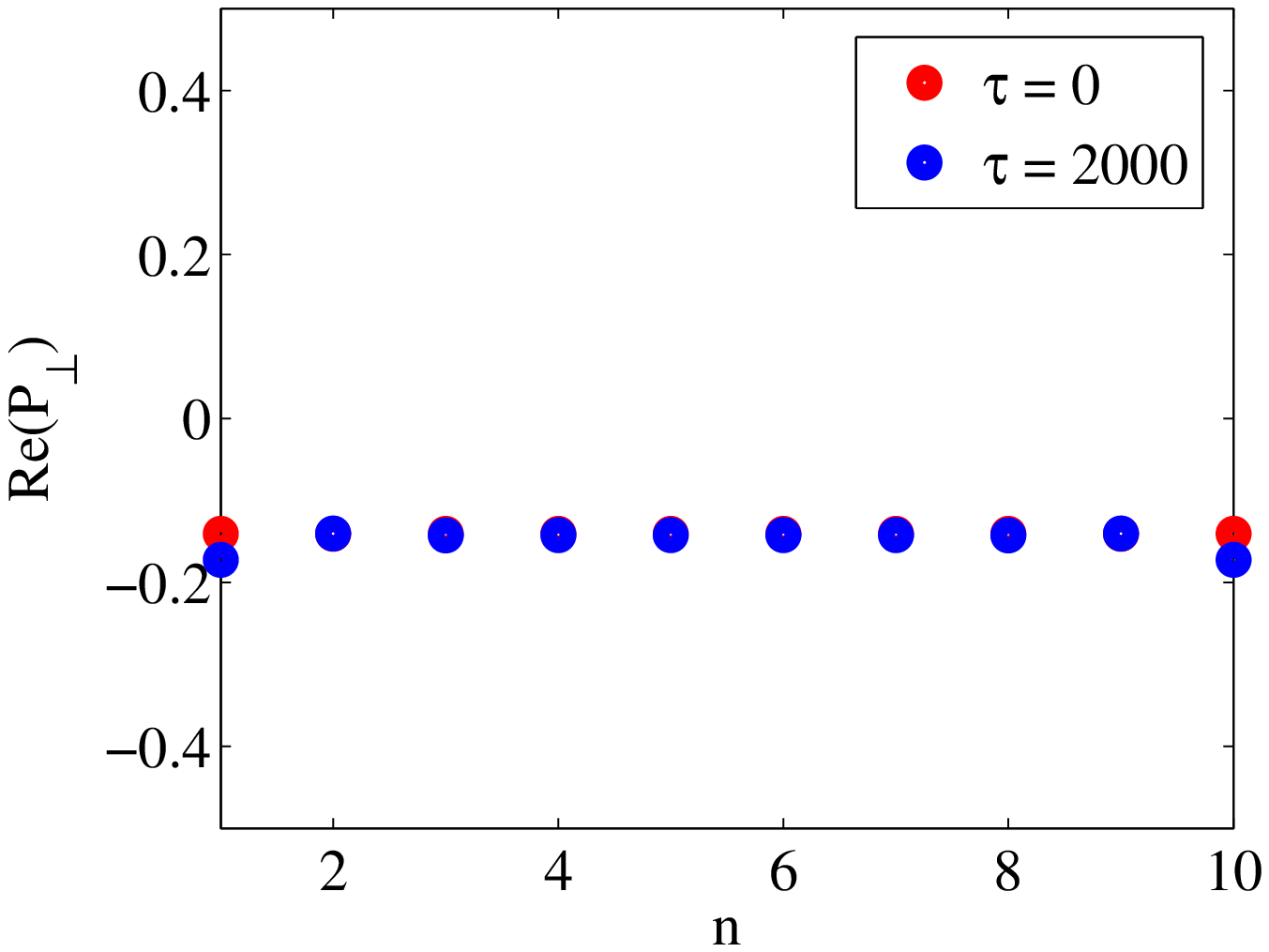}}\label{fig2e}
\subfigure[]{\includegraphics[width=5cm, height=4cm]{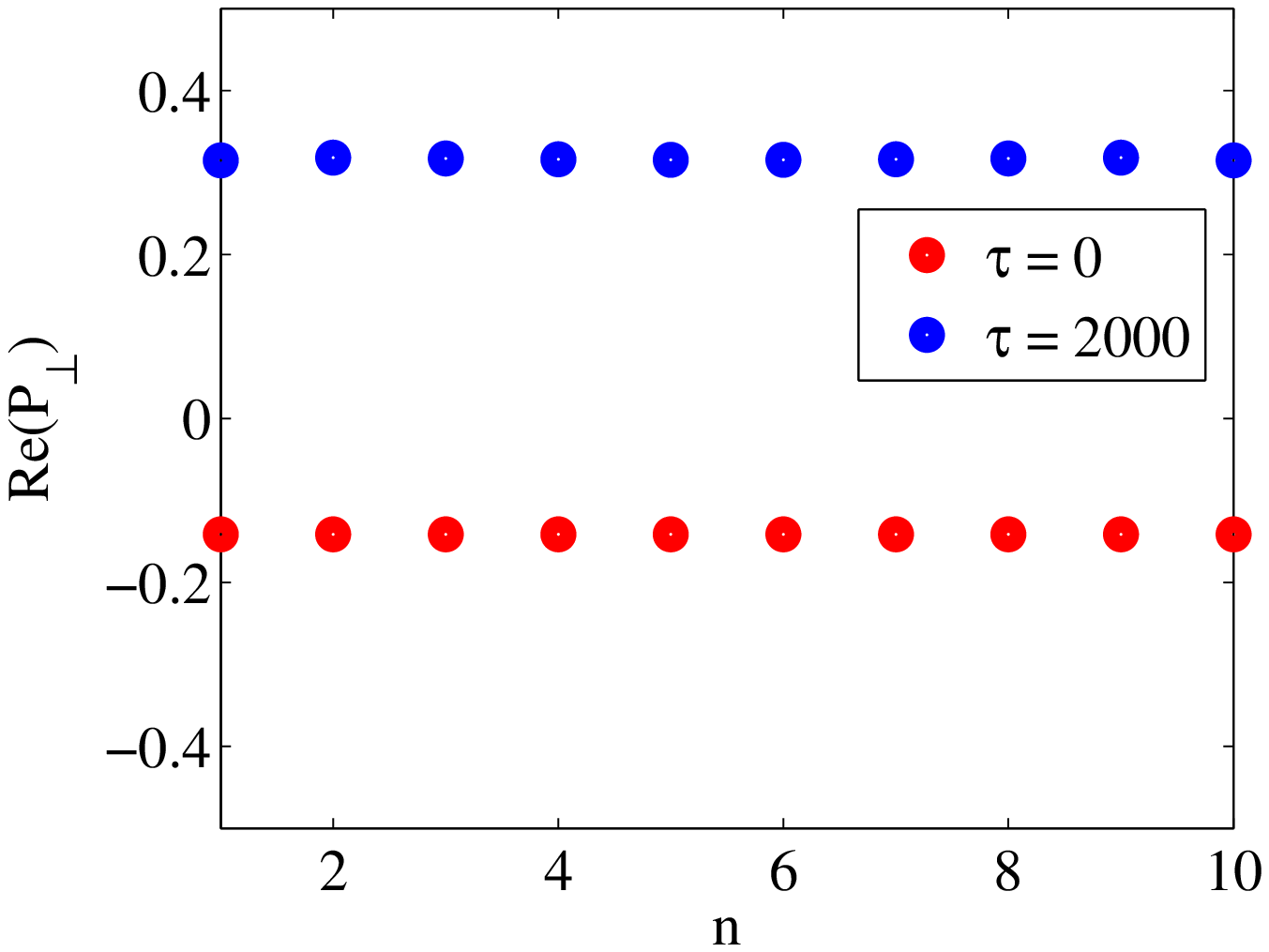}}\label{fig2f}
\caption{(Color online) (a) The bistability at $\Omega=-0.1$, (b) real parts of particle dipole moments with $\Omega=-0.1$, $|E_{0}|^2=1.9\times 10^{-4}$, (c) real parts of particle dipole moments with $\Omega=-0.1$, $|E_{0}|^2=2.0\times 10^{-4}$, (d) bistability at $\Omega=-0.2$, (e) real parts of particle dipole moments with $\Omega=-0.2$, $|E_{0}|^2=5.9\times 10^{-4}$, and (f) real parts of particle dipole moments with $\Omega=-0.2$, $|E_{0}|^2=6.0\times 10^{-4}$.}
\end{figure}

The emergence of a critical point in the low branch is important in the control of the localized optical energy in a finite number array. Thus, the influence of the emergence of the critical point on the environment parameter should be elucidated. The numerical simulations show that the emergence of the critical point is strongly dependent on the value of $\Omega$ as shown in Fig. 3 through the function curve, which separates the red and yellow areas versus $\Omega$.

\begin{figure}[htbp]
\centering\includegraphics[width=8cm,height=6cm]{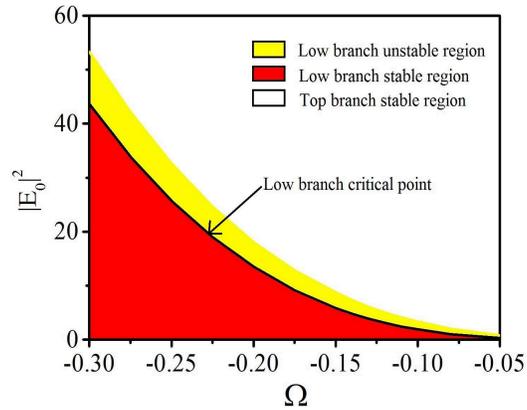}\label{fig3}
\caption{(Color online) Diagram of stability analysis with different $|E_{0}|^{2}$ and $\Omega$ values in the finite number of particles system. The central solid curve, which separates the red and yellow areas, is the function curve of the critical point. }
\end{figure}

The source of instability at the low branch in a finite particle system is analyzed.  A simulation of real-time evolution is given in Fig. 4(a), in which the initial point is the unstable solution from the low branch in the bistable curve. The red region refers to the solution of the particle dipole moment that remains in the low branch, and the yellow areas refer to the solution jump to the top branch, which are treated as the stable and unstable areas, respectively. From this panel, the instability of the system is produced from the system boundary. To study the relationship between such instability and the total number of particles, we introduce a character denoted as the unstable growth rate (UGR):
\begin{equation}
\mathrm{UGR}=\frac{n_{\mathrm{un}}}{n\cdot T}\times100\%
\end{equation}
where $n_{\mathrm{un}}$ is the number of unstable particles at $\tau=T$, $n$ is the total number of particles, and UGR represents the average growing number of unstable particles in unit of $\tau$. The UGRs as functions of $n$ with different values of $|E_{0}|^{2}$ are displayed in Fig. 4(b). From this panel, UGR has a larger value when $|E_{0}|^{2}$ is farther from the critical point in the unstable region in the low branch, as shown by the red curve. Although UGR$\neq0$, this is the essential for a finite system. But more importantly, when $n$ approaches infinity, UGR approaches zero, instability source from the edge effects can be neglected, the low branch can be treated as stable, which is similar to the previously reported result from an infinity system \cite{oe}.

\begin{figure}[htbp]
\centering
\subfigure[]{\includegraphics[width=8cm, height=6cm]{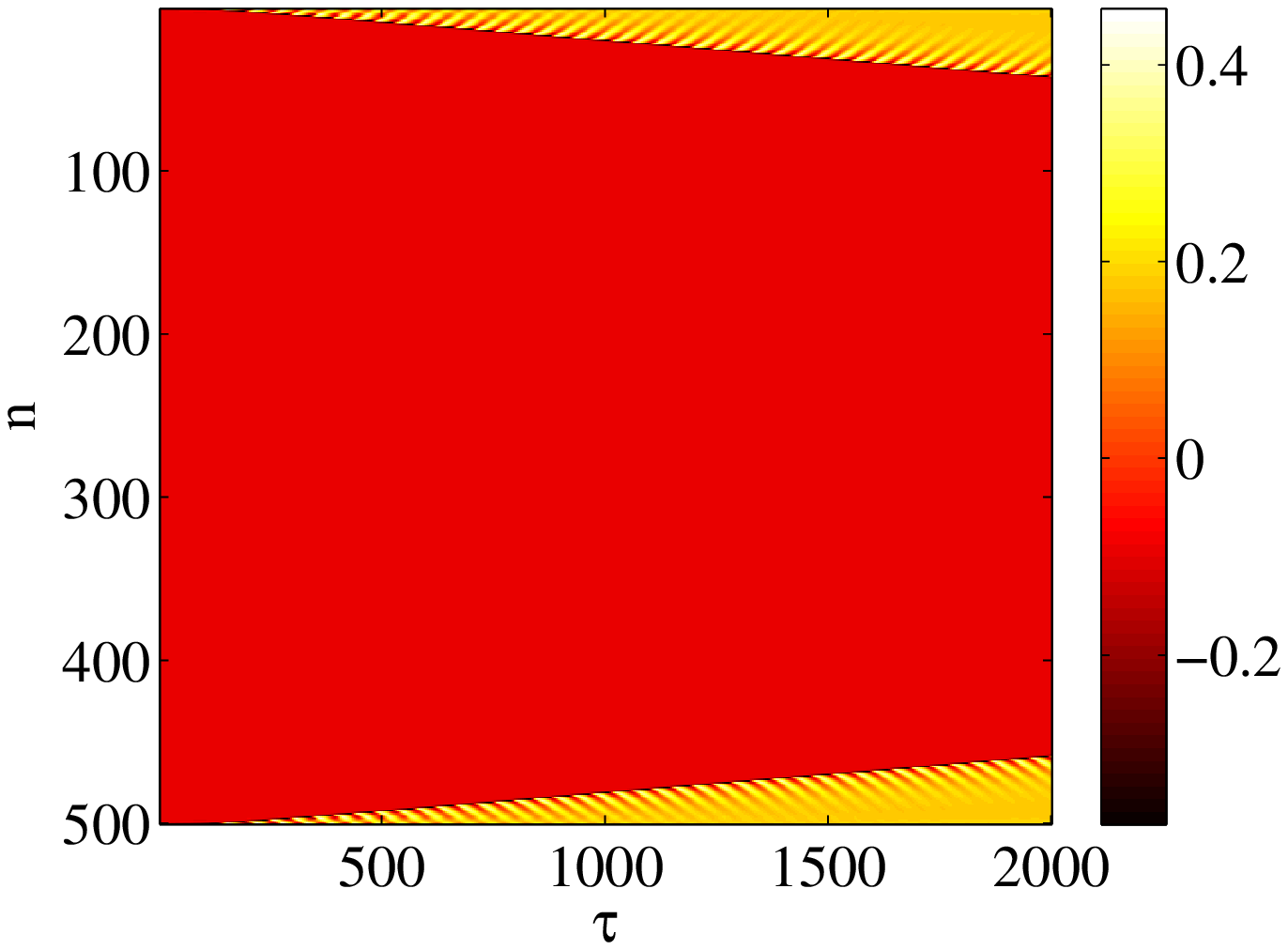}}\label{N500}
\subfigure[]{\includegraphics[width=8cm, height=6cm]{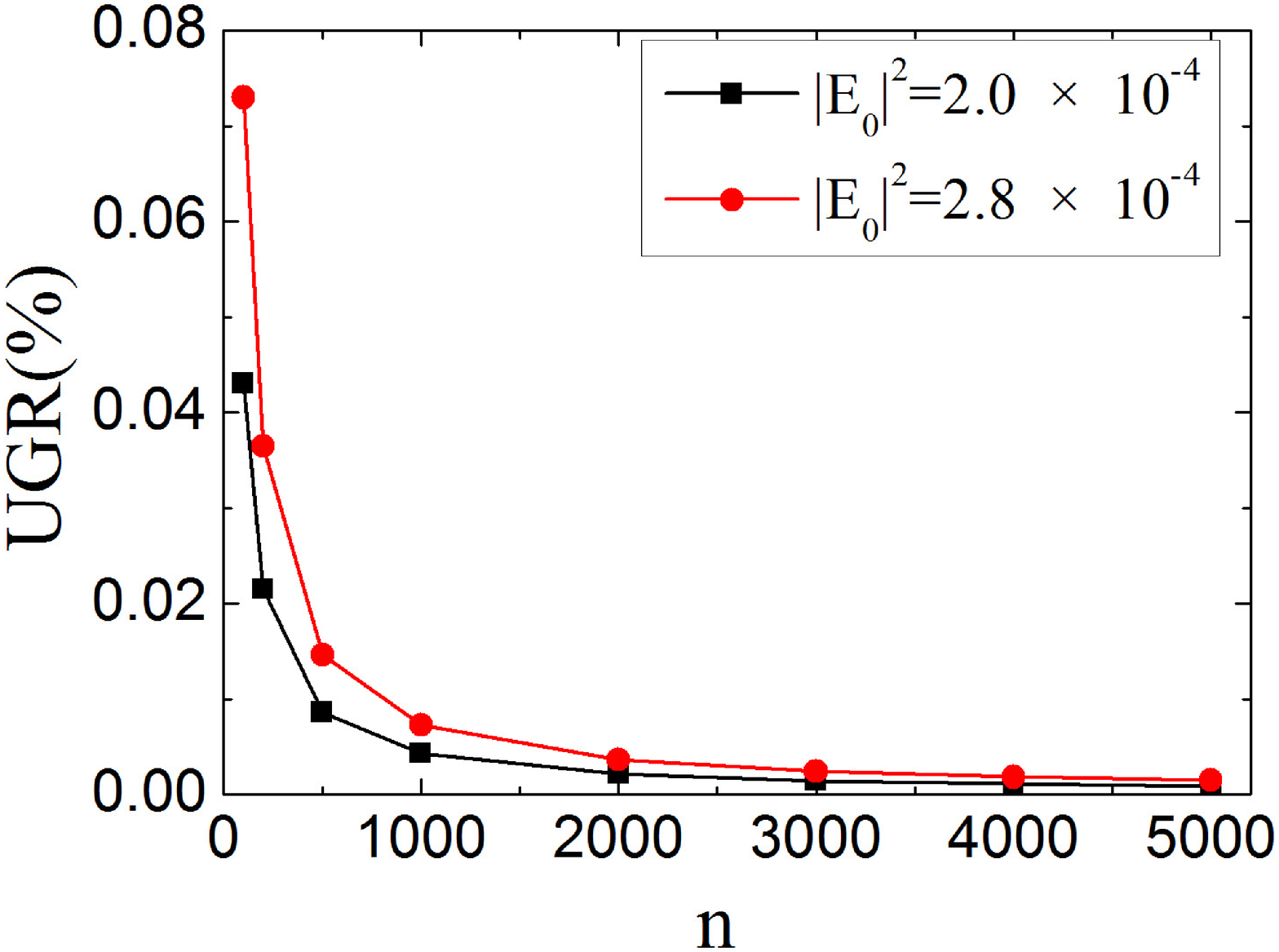}}\label{UGR}
\caption{(Color online) (a) Low branch real-time evolution of a finite system with 500 particles at $\Omega=-0.1$, $|E_{0}|^2=2.0\times 10^{-4}$, in which the red area refers to stable particles, and the other colored regions denote unstable particles, (b) UGR versus $n$ measured at $T=2000$.}
\end{figure}

\section{Control the formation of dark soliton in finite number of particles array}

The interpretation of solitons as a pair of tightly coupled kinks has been widely exploited for many similar systems \cite{clerc, ishimori, peyard, karpan}. In these systems, although the edge effect resulted in the instability was a well-known physical phenomena, how this effect affects the formation of dark dipole solitons have not been reported both in infinite and finite systems. In our work, we not only demonstrate how to form a dipole siliton by considering the edge effects, but also find the extremely narrow width soliton (just a few tens of nanometer) in a finite systems.

Dipole kinks (switching wave) can be constructed in terms of such stability analysis on the basis of the bistability of dipole moments. For instance, as shown in Fig. 5(a), a dipole kink can be formed by setting half of the particles at a stable region at the low branch and the other half at the top branch. The dynamical evolution of such kink was investigated by Ref. \cite{oe}. Using a similar strategy, two oppositely polarized kinks is constructed [Fig. 5(b)], which can be used to form a dark dipole soliton. Although in Ref. \cite{oe,sr}, Noskov et al. showed typical examples of dark soliton, they did not provide a detailed discussion on how to control the fomation of the dark solitons. Additionally,  they did not optimize their results, e.g., the narrowest dark soliton. In this section, we will discuss in detail how to control the formation of dark solitons and obtain an optimized soliton in this system.
\begin{figure}[htbp]
\centering
\subfigure[]{\includegraphics[width=8cm, height=6cm]{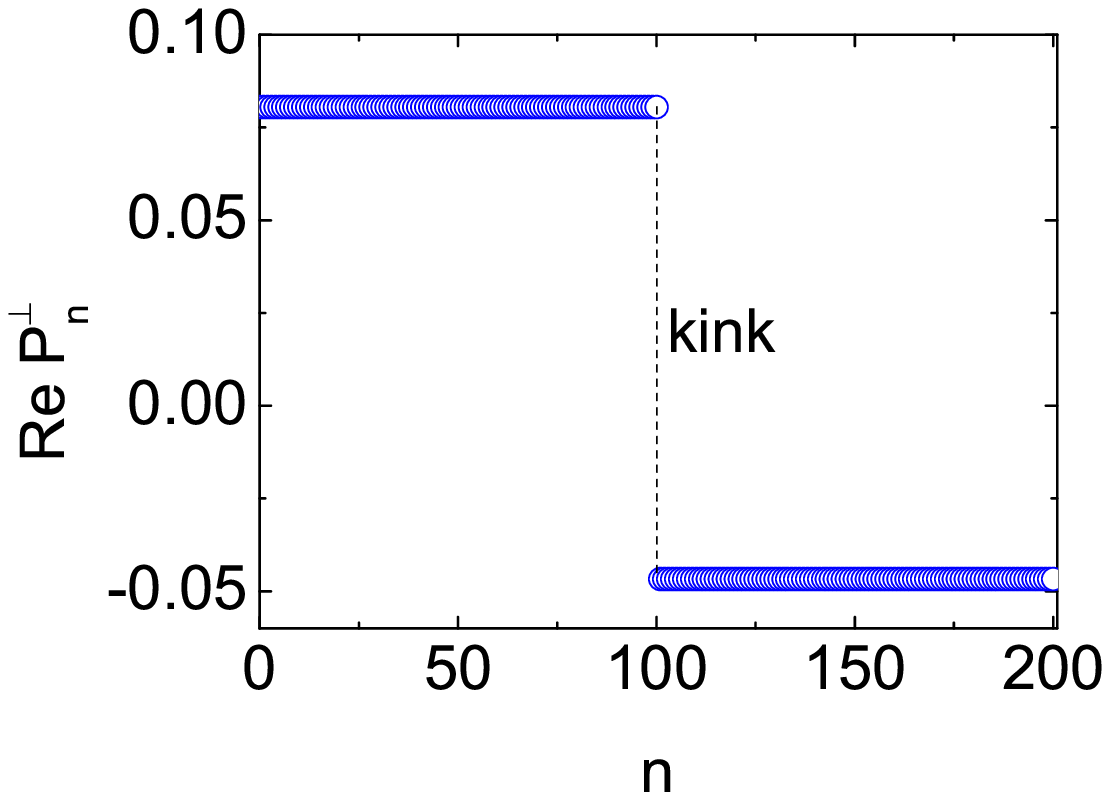}}\label{kink1}
\subfigure[]{\includegraphics[width=8cm, height=6cm]{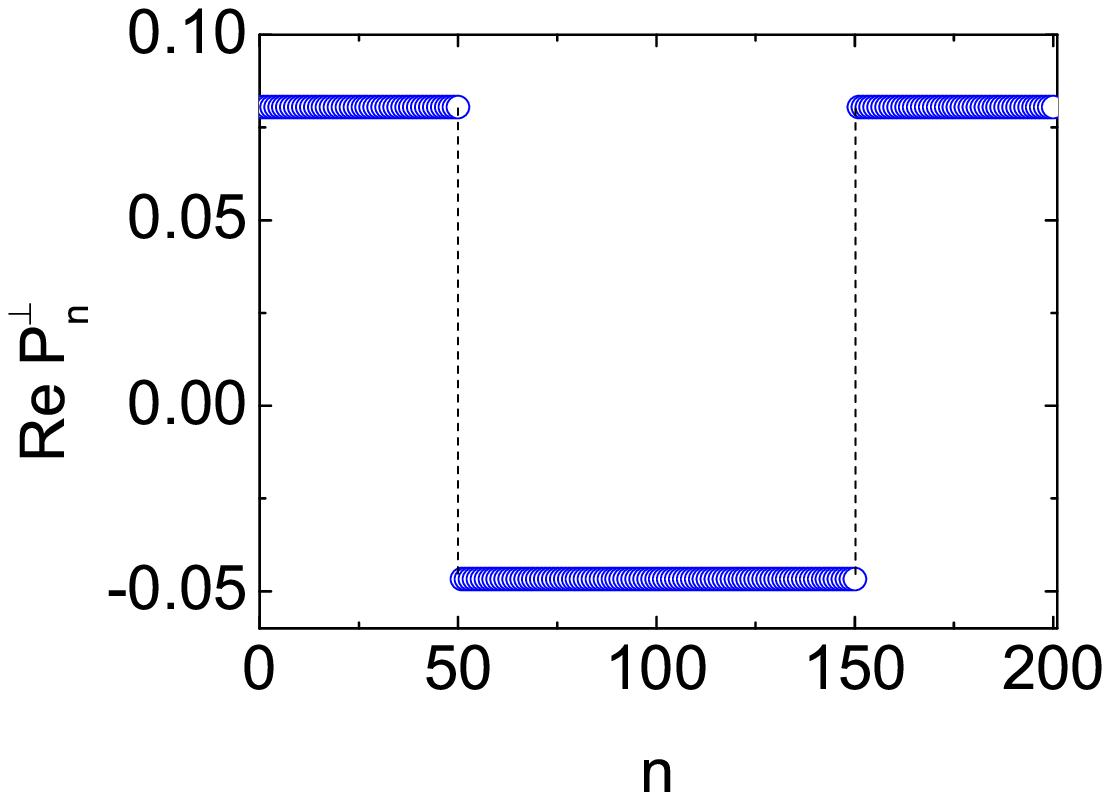}}\label{kink2}
\caption{(Color online) Generation of kinks in a 200-particle system excited by the homogeneous electric field of $|E_{0}|^2=0.60\times10^{-4}$ under the $\Omega=-0.1$ at $\tau=0$ (a) one kink and (b) two kinks, which is selected as the initial condition in the simulation to form a dark soliton.}
\end{figure}

Two essential factors should be considered in the formation of dark solitons. One factor is the external field intensity (i.e., $|E_{0}|^{2}$), and the other is the initial kink-kink width. In this study, we focus on the influence of the external electric field and fixed initial kink-kink width on the symmetrical structure, as shown in Fig. 5(b). Under this condition, with an appropriate intensity of the external field, the two kinks start to move toward each other with similar speeds, which become zero prior to the collision, resulting in the formation of a dark dipole soliton (DDS). Typical examples of such dynamical process with different values of $|E_{0}|^{2}$ are displayed in Fig. 6, in which the red region refers to the solution of particle dipole moment that remains in the low branch of the bistable curve, and the yellow region refers to the top branch. The numerical result shows the presence of a threshold $|E_{\mathrm{f}}|^{2}$. When $|E_{0}|^{2}<|E_{\mathrm{f}}|^{2}$, the kinks do not move, resulting in the width of the dark soliton to be equal to that of the initial condition [Fig. 6a]. By contrast, when $|E_{0}|^{2}>|E_{\mathrm{f}}|^{2}$, the kinks start to move toward each other and then halt with a stable width between each other [Figs. 6(b) and 6(c)]. The dynamics of the kink from stationary to motion against the magnitude of the intensity of the external field is studied by Ref. \cite{oe}. However, another threshold exists, namely $|E_{\mathrm{v}}|^{2}$, is larger than the former. The kinks collide with each other, causing close up the formation of dark solitons when $|E_{0}|^{2}>|E_{\mathrm{v}}|^{2}$ [Fig. 6(d)]. Therefore, the width of the dark soliton can be controlled by setting the $|E_{0}|^{2}$ into the interval of $\left[|E_{\mathrm{f}}|^{2}, |E_{\mathrm{v}}|^{2}\right]$. The simulations show that the position and the width of the interval in the domain of $|E_{0}|^{2}$ can be controlled by detuning $\Omega$. For example, in Fig. 7(a), the position of the interval is approximately $0.57\times10^{-4}$ to $0.60\times10^{-4}$, and its width is $0.03\times10^{-4}$ for $\Omega=-0.1$. By contrast, in Fig. 7(b), the position of the interval shifts $8.3\times10^{-4}$ to $8.4\times10^{-4}$, and the width increases to $0.1\times10^{-4}$. Fig. 7(c) displays the right border of the interval, i.e., $|E_{v}|^{2}$ as a function of $\Omega$, which shows that this threshold decays as $\Omega$ increases to the resonant point.

\begin{figure}[htbp]
\centering
\subfigure[]{\includegraphics[width=8cm, height=6cm]{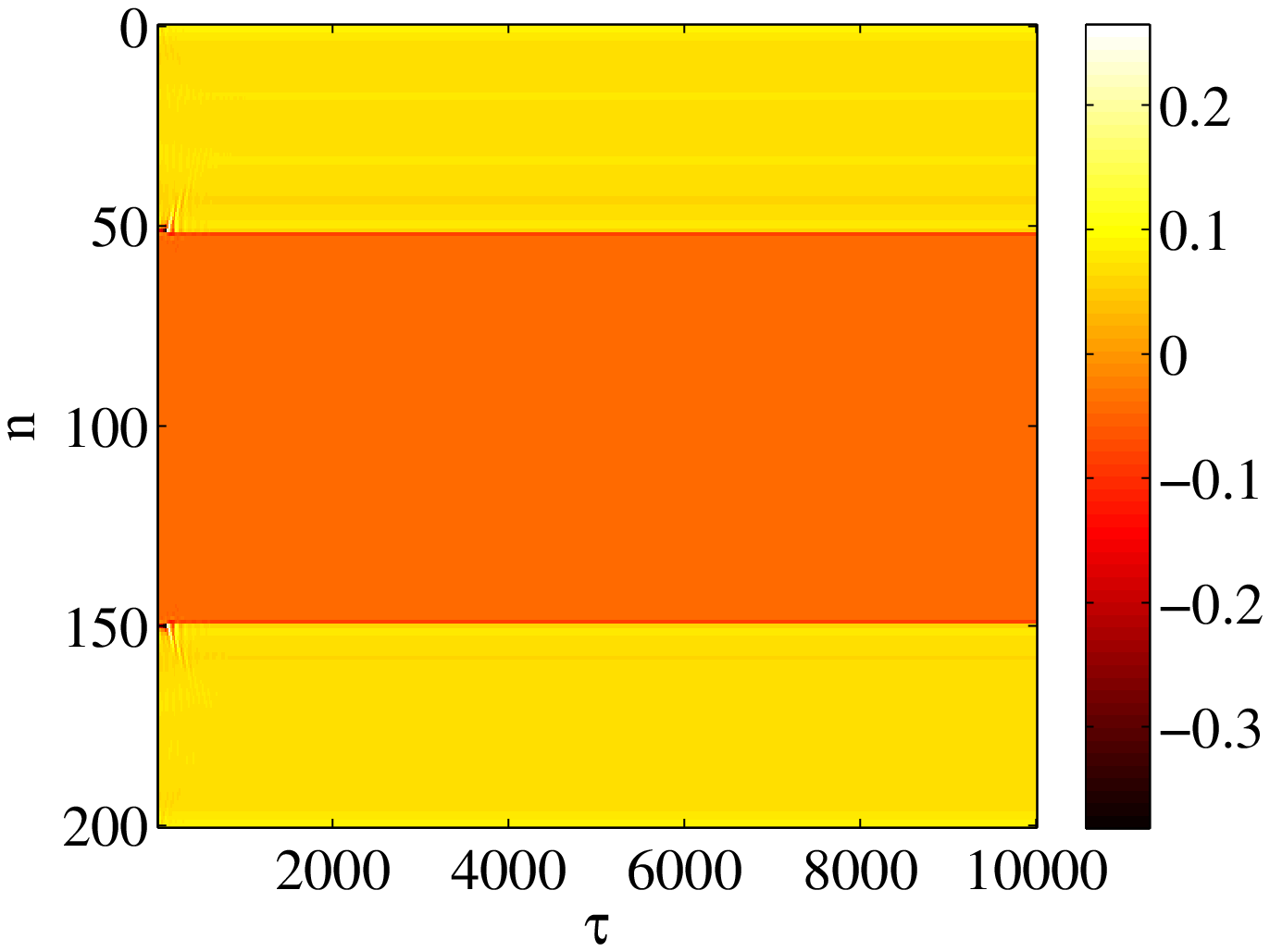}}
\subfigure[]{\includegraphics[width=8cm, height=6cm]{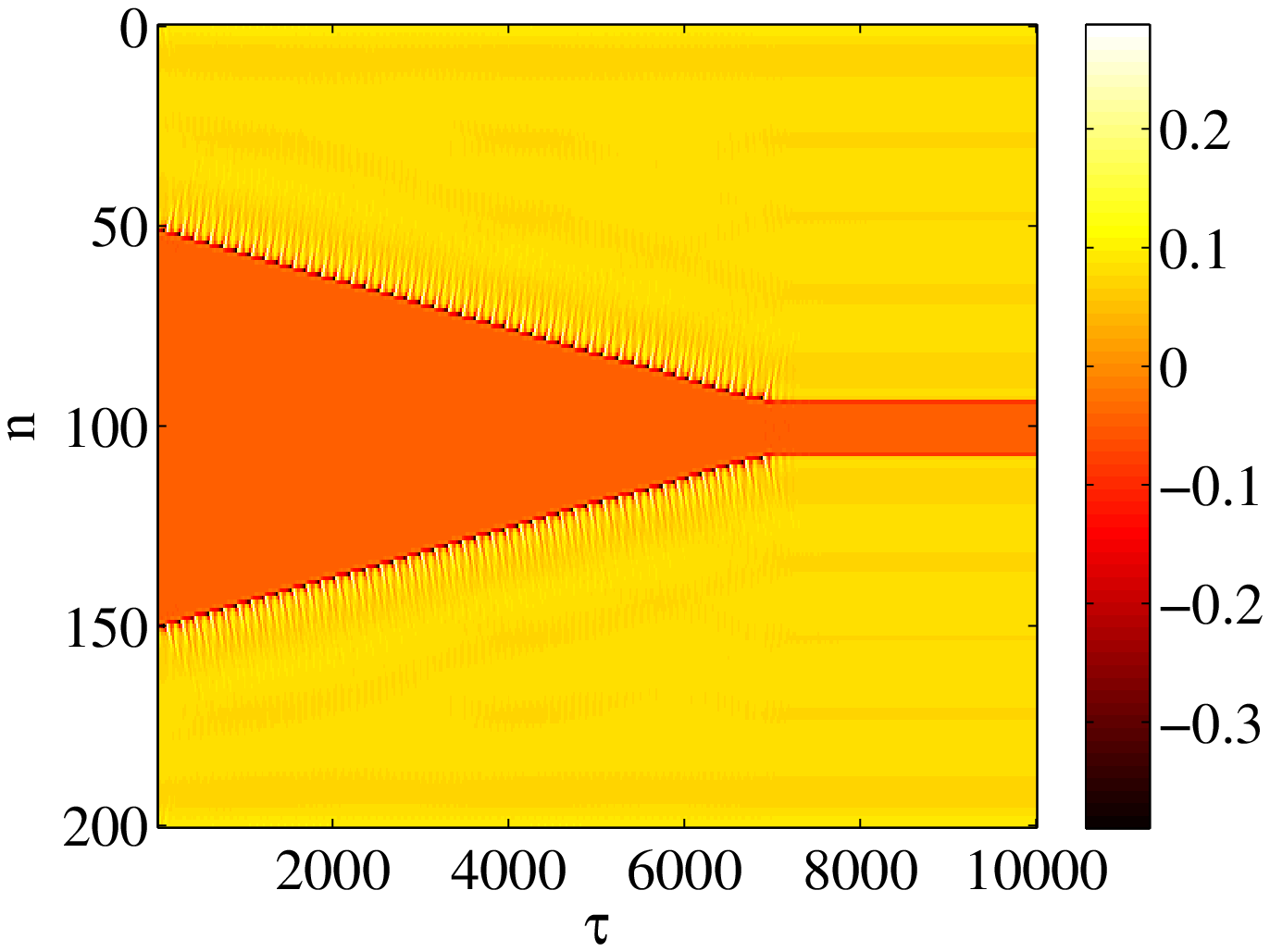}} \\
\subfigure[]{\includegraphics[width=8cm, height=6cm]{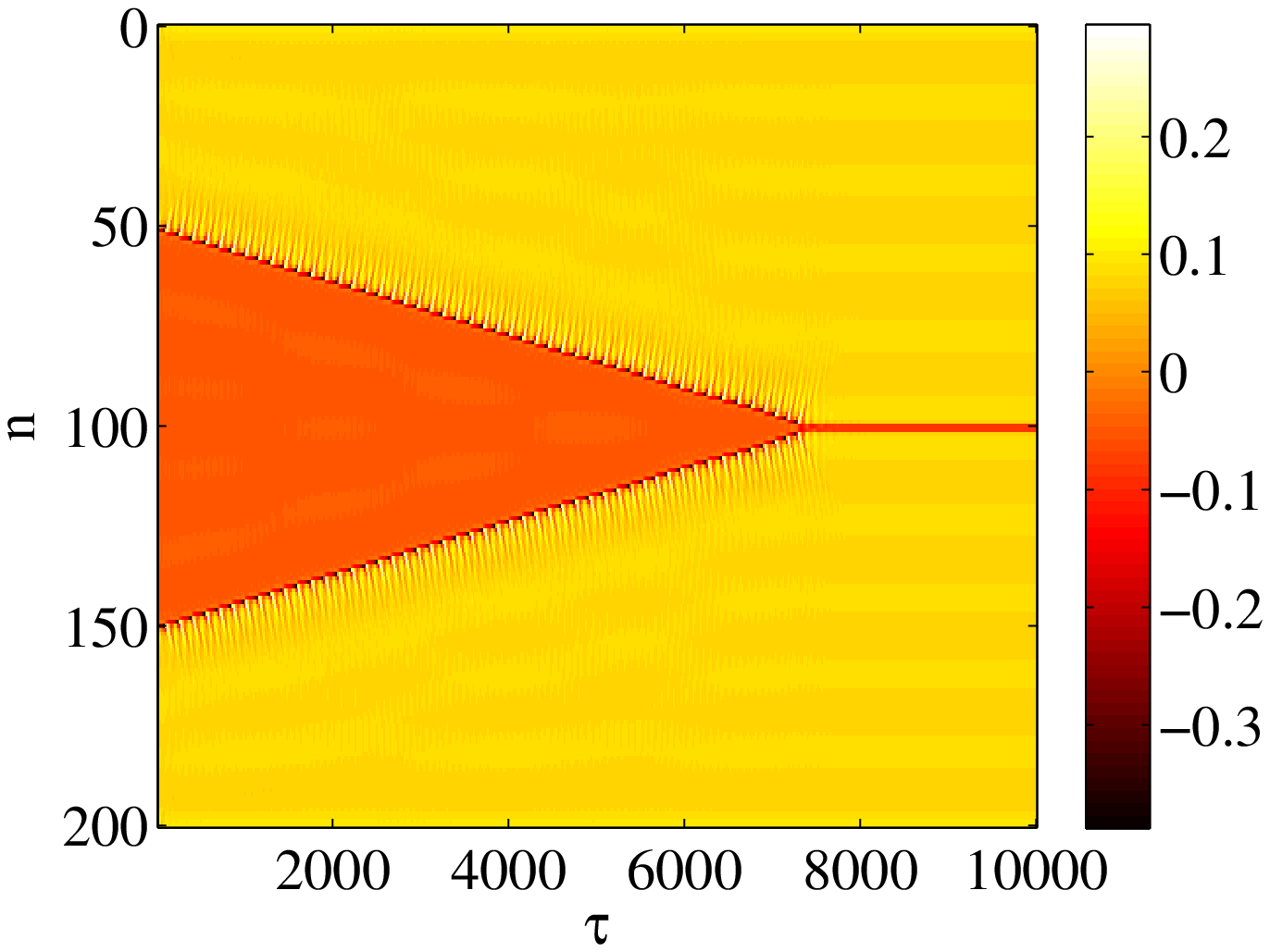}}
\subfigure[]{\includegraphics[width=8cm, height=6cm]{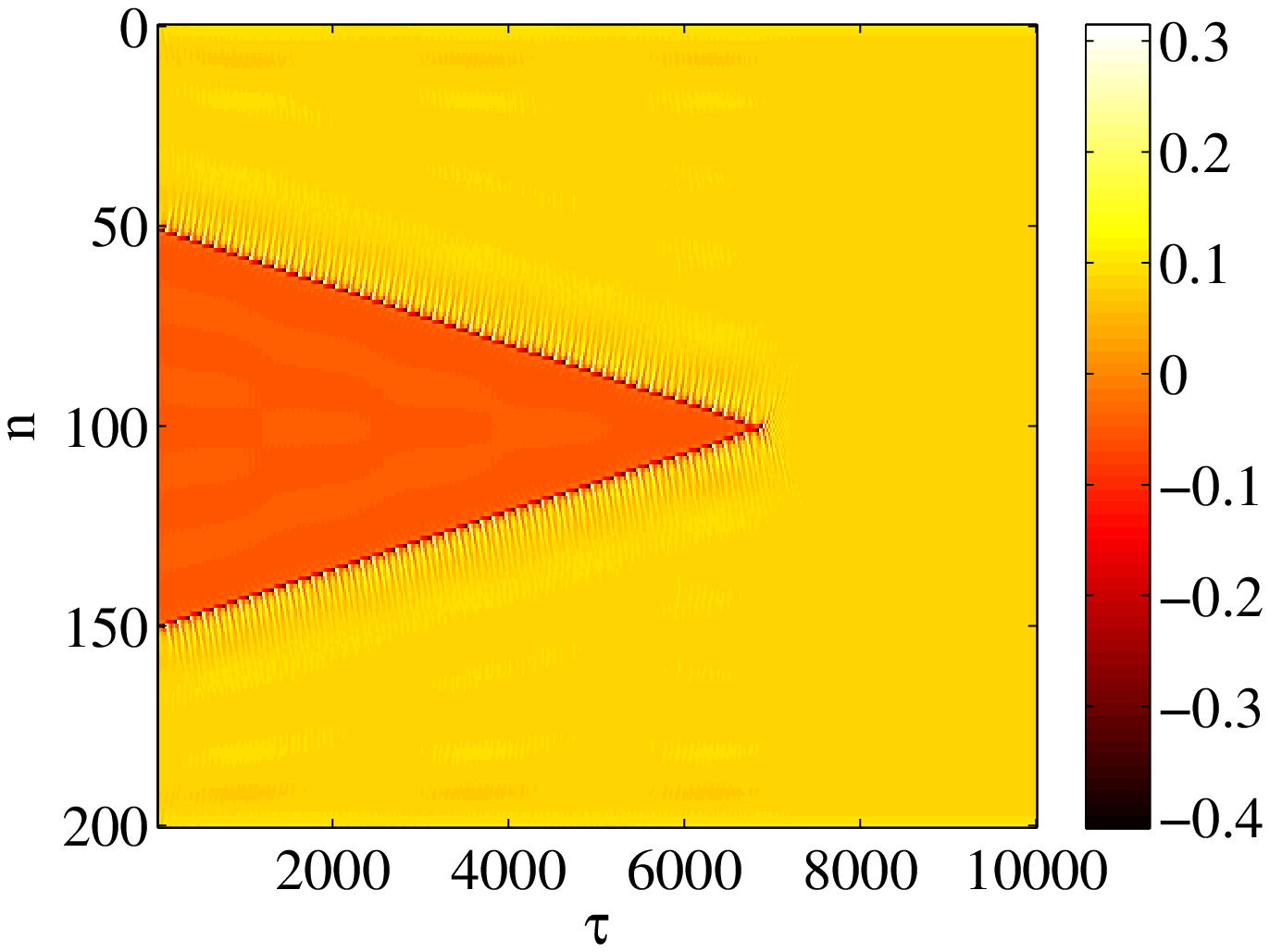}}
\caption{(Color online) Tailoring the width of a dark soliton by modulating the external field under the $\Omega=-0.1$, (a) $|E_{0}|^2=0.55\times10^{-4}$, (b) $|E_{0}|^2=0.59\times10^{-4}$, (c) $|E_{0}|^2=0.60\times10^{-4}$, (d) $|E_{0}|^2=0.62\times10^{-4}$.  The total number of particles in the system is 200.}
\end{figure}

\begin{figure}[htbp]
\centering
\subfigure[]{\includegraphics[width=6cm, height=5cm]{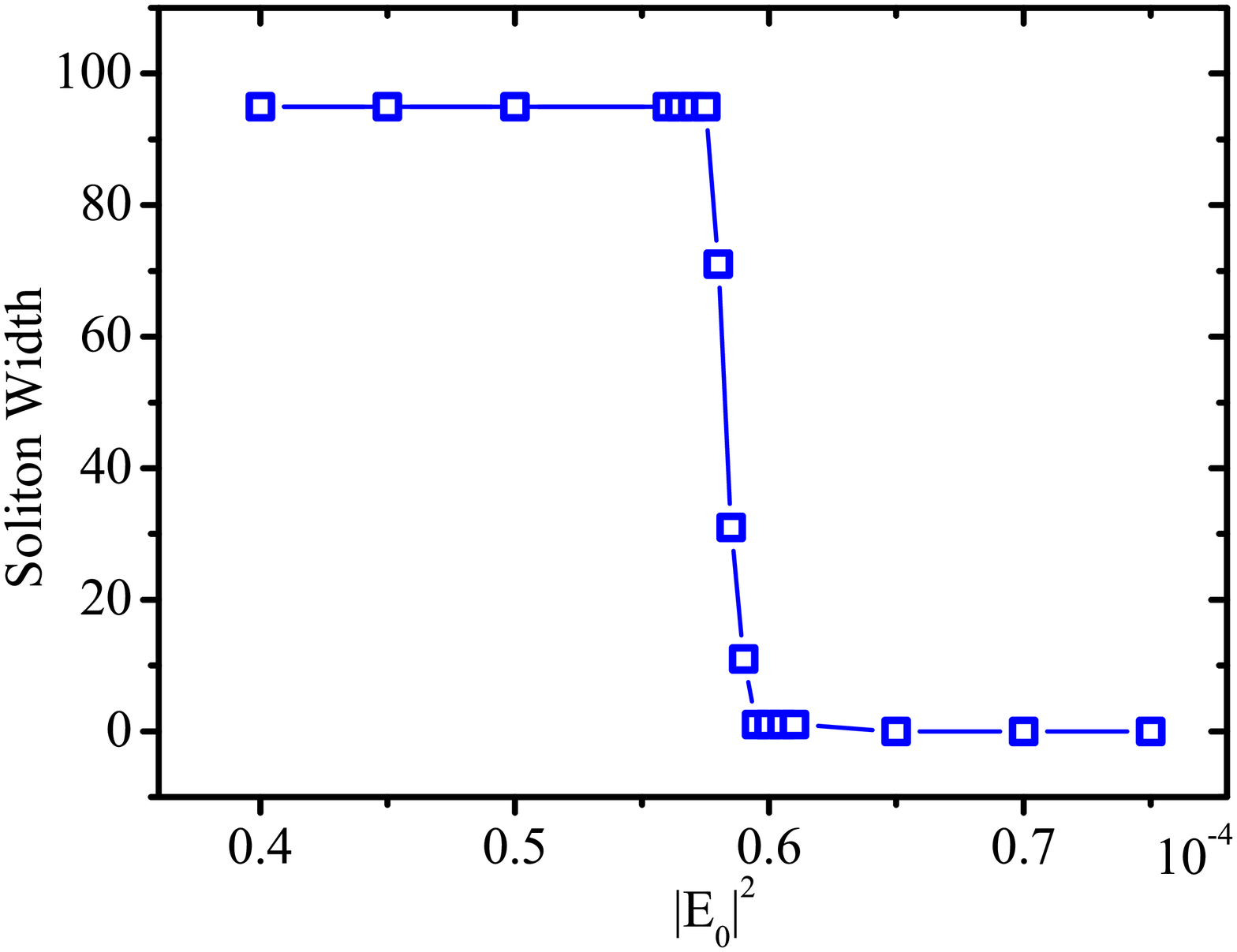}}
\subfigure[]{\includegraphics[width=6cm, height=5cm]{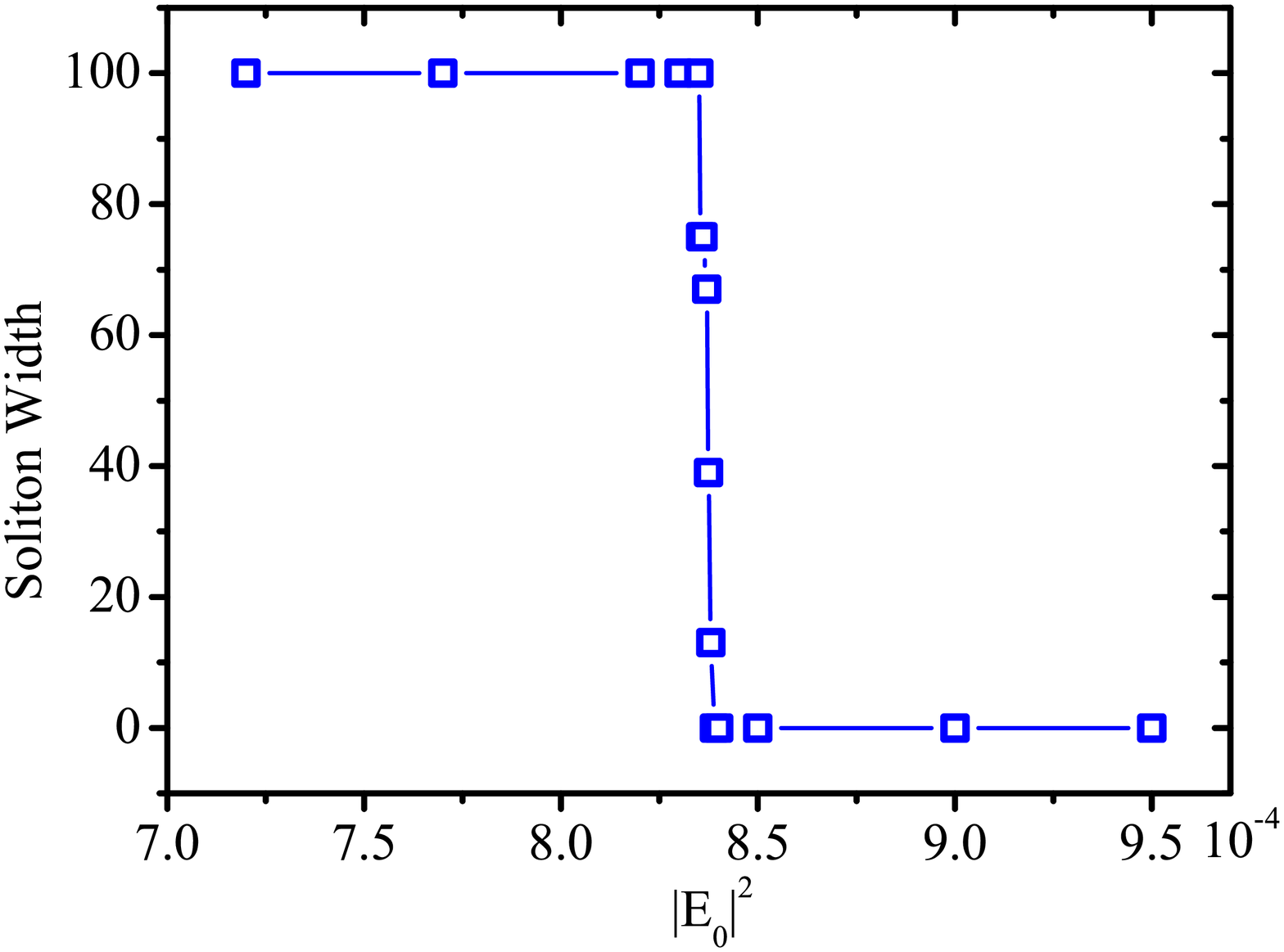}}
\subfigure[]{\includegraphics[width=6cm, height=5cm]{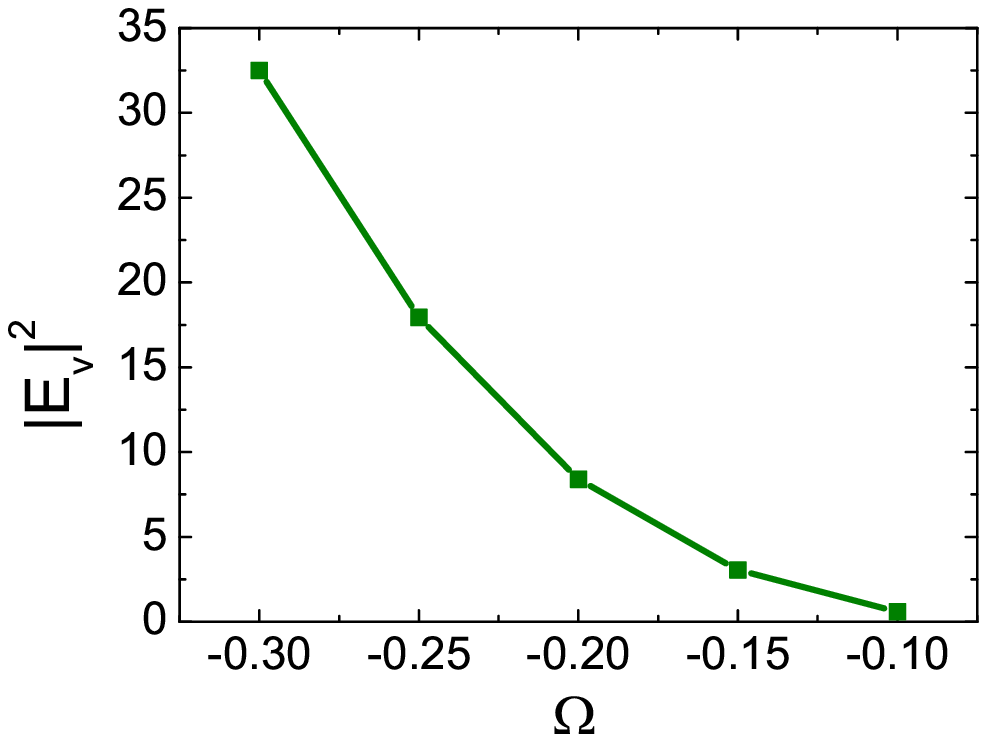}}
\caption{(Color online) Relationship between the width of a dark soliton and the external field intensity. (a) $\Omega=-0.1$, (b) $\Omega=-0.15$, (c) The vanishing points for dark soliton with different values of $|E_{\mathrm{v}}|^{2}$ and $\Omega$. The total number of particles in the system is 200 and the initial kink-kink widths is 100 particles.}
\end{figure}
Typically, further simulations show the presence of a special domain in the plane of $(|E_{0}|^{2},\Omega)$ [blue area in Fig. 8(a)], which is attached to the border of $|E_{\mathrm{v}}(\Omega)|^{2}$ and has a double monopole dark soliton (DMDS) inside. A typical example of the DMDS is displayed in Fig. 8(b). This soliton contains only two stable particles at the low branch and has a width of 30 nm (the distance between the adjacent particles in the scheme). This width is far smaller than the previous result in Ref. \cite{oe}. In Fig. 8(a),  in a 200-particle system, the DMDS is supported by the detuning frequency, which ranges from $-0.08$ to -0.11(corresponding to the optical wavelength from 435 nm to 450 nm), and the external intensity $|E_{0}|^2$ from $0.28\times10^{-4}$ to $0.88\times10^{-4}$ (approximately corresponding to a laser intensity from 29 MW$/$cm$^2$ to 92 MW$/$cm$^2$), and the vertical width of the domain (blue area) is approximately $0.02\times10^{-4}$, corresponding to a laser intensity of 2 MW$/$cm$^2$. These values are all reasonable in a real experiment estimation.

\begin{figure}[htbp]
\centering
\subfigure[]{\includegraphics[width=8cm, height=6cm]{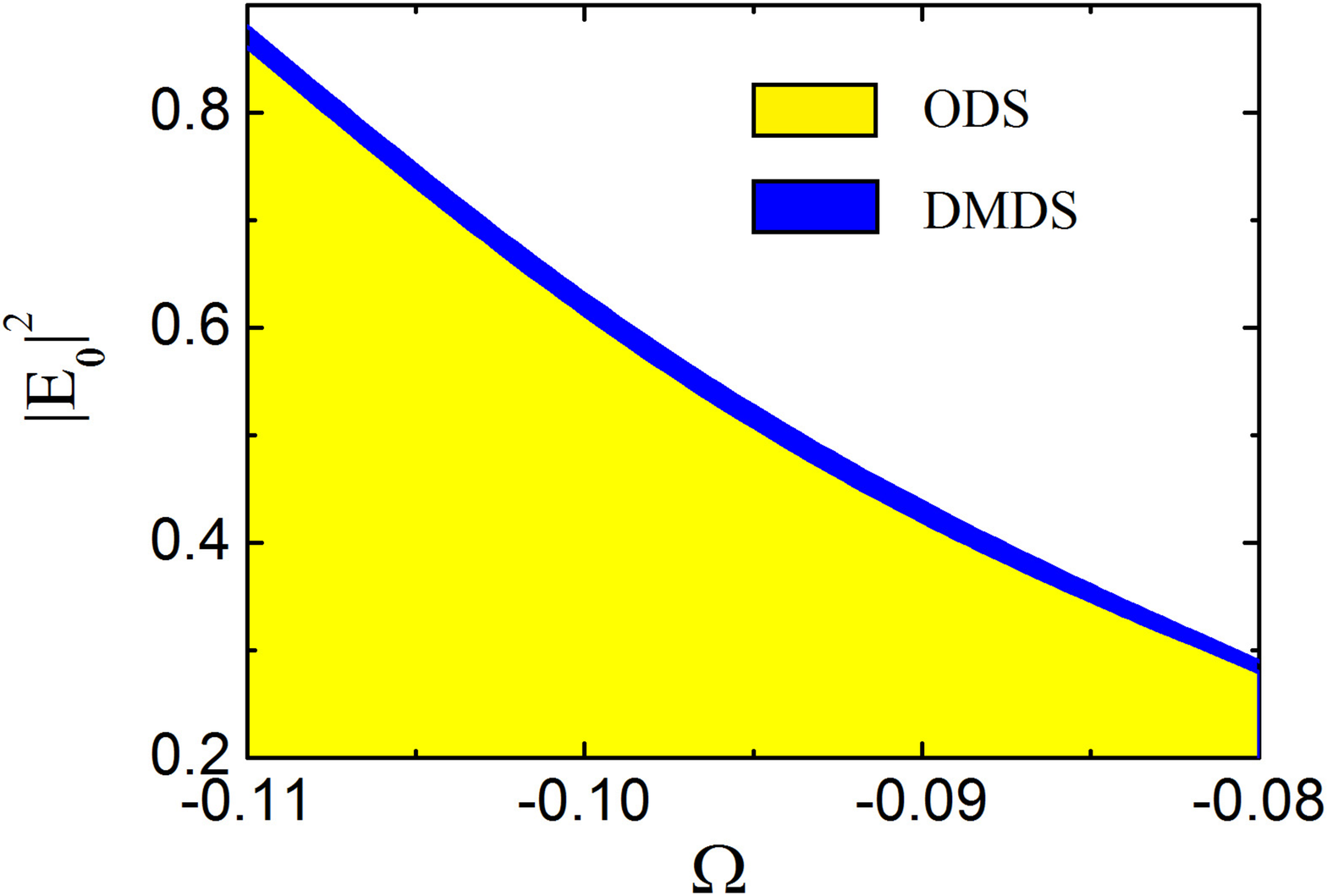}}
\subfigure[]{\includegraphics[width=8cm, height=6cm]{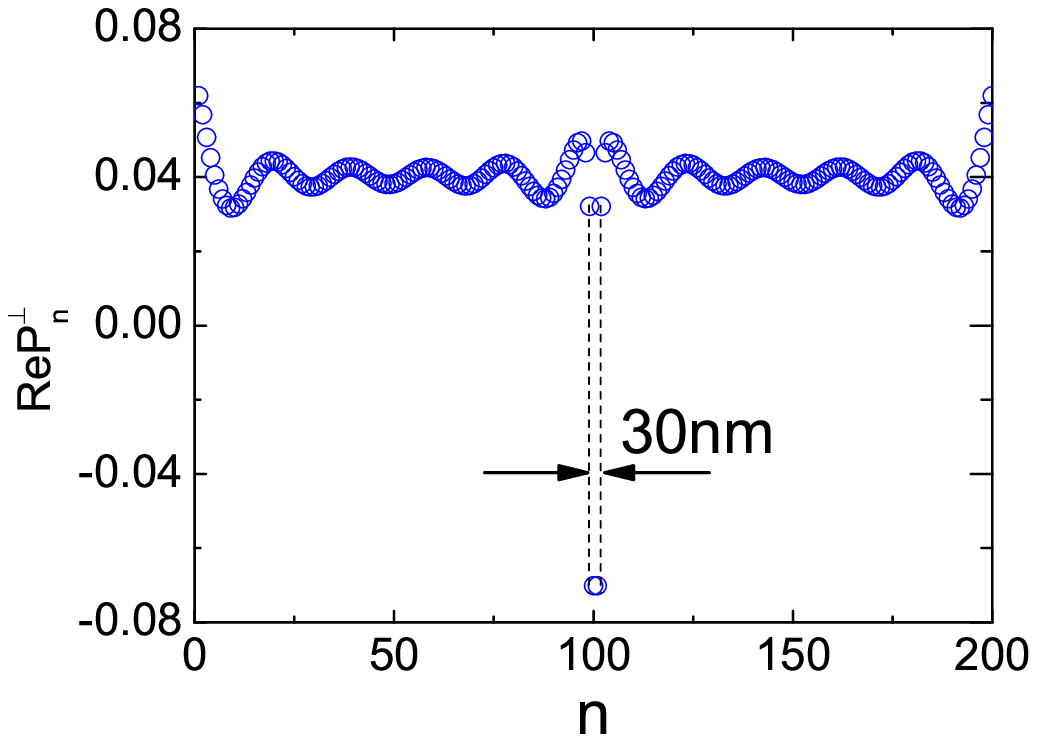}}
\caption{(Color online) DMDS. (a)Formation of DMDS in the blue zone. The yellow zone refers to the ordinary dark solitons (ODS), which contain more than two particles at the low branch, and the blank zone refers to the absence of a dark soliton. The total number of particles in the system is 200; (b)DMDS with $\Omega=-0.09$, $|E_{0}|^2$=0.41$\times$ 10$^{-4}$, initial kink-kink width of 100 particles.}
\end{figure}

\section{Conclusion}
We analyzed the stabilities of the dipolar mode in a finite number nanoparticle array, and the formation of controllable width of dark soliton in the same system. In this study, we performed numerical analysis of a silver array with finite number of nanoparticles that are excited by a homogenous electric field with the transversal polarization with consideration of the linearly dipole-dipole interaction and cubic nonlinearity. We studied the formation mechanism of dark soliton in the stationary states, analyzed the stability of the dipole moments in this finite number nanoparticle system and demonstrated the tailoring of the width of a dark soliton by changing the initial external electric field at different detuning frequencies. Moreover, we found DMDS consisting of two particles with the width of approximately $30nm$. These results may have potential applications in the subwavelength highly precise detection.\\

\begin{acknowledgments}
The authors especially thank Prof. Chongjun Jin for his time and efforts on this work and allowing to conduct the results in his labs.\\
This work is supported by the National Natural Science Foundation of China (Grant Nos.11104083, 11204089, 61172011) and the Guangdong Natural Science Foundation (Grant No. 10151064201000006). We would like to express our gratitude to the Modern Educational Technology Center of South China Agricultural University for giving us access to its computing facility.
\end{acknowledgments}

\appendix*

\section{parameters}
The radius and the distance between the particles are fixed to $a=10$ nm and $d=30$ nm, respectively. An external optical field $E^{\mathrm{ex}}$ is launched into the host and stimulates the particles. The dispersion of the host can be neglected because  the permittivity of SiO$_{2}$ is nearly $\varepsilon_{h}\simeq2.15$ for the optical wavelength range.The linear part of the dielectric constan of the silver particle follows the Drude model, which can be expressed as $\varepsilon_{\mathrm{Ag}}^{\mathrm{L}}=\varepsilon_{\infty}-\omega_{P}^{2}/[\omega(\omega-i\nu)]$, where $\varepsilon_{\infty}=4.96$, $\hbar\omega_{P}=9.54$ eV and $\hbar\nu$=0.55 eV in this study \cite{oe18}. The nonlinear part of the dielectric constant is selected as the standard cubic type, which can be obtained as $\varepsilon_{\mathrm{Ag}}^{\mathrm{NL}}=\chi^{(3)}|E_{n}|^{2}$, where $\chi^{(3)}\simeq3\times10^{-9}$ esu for the silver spheres with 10nm radius\cite{oe20}, and $E_{n}$ is the local field inside the $n^{\mathrm{th}}$ particle. The frequency of the surface plasmon resonance of the silver nanoparticles, namely $\omega_{0}$, which can be expressed as $\omega_{0}=\omega_{P}/\sqrt{\varepsilon_{\infty}+2\varepsilon_{h}}$.

%

\bibliographystyle{plain}
\bibliography{apssamp}

\begin{thebibliography}{99}
\bibitem{Maier}S. A. Maier, \emph{Plasmonics: fundamentals and applications}. (Springer, 2007).

\bibitem{Tliu}T. Liu, Y. Shen, Q. Zhu, Z. Zhou, and C. Jin, JOSA B, \textbf{30}, 1420.(2013).
\bibitem{shenyang1}Y. Shen, X. Chen, Z. Dou, N. P. Johnson, Z. K. Zhou, X. Wang, and C. Jin, Nanoscale, \textbf{4}, 5576.(2012).
\bibitem{shenyang2}Y. Shen, M. Liu, J. Li, X. Chen, H. X. Xu, Q. Zhu, X. Wang, and C. Jin, Plasmonics, \textbf{7}, 221.(2012).
\bibitem{shenyang3}Y. Shen, M. Liu, Q. Wang, P. Zhan, Z. Wang, Q. Zhu, X. Chen, S. Jiang, X. Wang, and C. Jin, Nanoscale, \textbf{4}, 2255.(2012).
\bibitem{sr11}D. K. Gramotnev, and S. I. Bozhevolnyi, Nat. Photonics, \textbf{4}, 83.(2010).
\bibitem{sr10}M. Pelton, J. Aizpurua, and G. Bryant, Laser Photonics Rev., \textbf{2}, 136.(2008).

\bibitem{belov1}A. Al\`{u}, P. A. Belov, and N. Engheta, New J. Phys., \textbf{13}, 033026.(2011).
\bibitem{prl3a}M. Quinten, A. Leitner, J. R. Krenn, and F. R. Aussenegg, Opt. Lett., \textbf{23}, 1331.(1998).
\bibitem{prl3b}K. Li, M. I. Stockman, and D. J. Bergman, Phys. Rev. Lett., \textbf{91}, 227402.(2003).
\bibitem{oe4}J. Takahara, S. Yamagishi, H. Taki, A. Morimoto, and T. Kobayashi, Opt. Lett., \textbf{22}, 475.(1997).
\bibitem{belov2} A. Al\`{u}, P. A. Belov, and N. Engheta, Phys. Rev. B, \textbf{80}, 113101.(2009).

\bibitem{martti}M. Kauranen, and A. V. Zayats, Nat. Photonics, \textbf{6}, 737.(2012).

\bibitem{sr15}N. N. Lepeshkin, A. Schweinsberg, G. Piredda, R. S. Bennink, and R. W. Boyd, Phys. Rev. Lett., \textbf{93}, 123902.(2004).

\bibitem{sr16}N. C. Panoiu, and R. M. Osgood, Nano Lett., \textbf{4}, 2427.(2004).

\bibitem{sr17}W. Fan, S. Zhang, N. C. Panoiu, A. Abdenour, S. Krishna, R. M. Osgood, Jr., K. J. Malloy, and S. R. J. Brueck, Nano Lett., \textbf{6}, 1027.(2006).
\bibitem{sr18}J. A. H. Van Nieuwstadt, M. Sandtke, R. H. Harmsen, F. B. Segerink, J. C. Prangsma, S. Enoch, and L. Kuipers, Phys. Rev. Lett., \textbf{97}, 146102.(2006).
\bibitem{sr19}M. W. Klein, M. Wegener, N. Feth, and S. Linden, Opt. Express, \textbf{15}, 5238.(2007).
\bibitem{sr20}Y. Zhang, N. K. Grady, C. Ayala-Orozco, and N. J. Halas, Nano Lett., \textbf{11}, 5519.(2011).
\bibitem{sr21}J. Butet, I. Russier-Antoine, C. Jonin, N. Lascoux, E. Benichou, and P. F. Brevet, Nano Lett., \textbf{12}, 1697.(2012).
\bibitem{noskov2012}R. E. Noskov, A. E. Krasnok, and Y. S. Kivshar, New J. Phys., \textbf{14}, 093005.(2012).
\bibitem{lapshina}N. S. Lapshina, R. E. Noskov, and Y. S. Kivshar, JETP Lett., \textbf{96}, 759.(2013).


\bibitem{prl4a}Y. Liu, G. Bartal, D. A. Genov, and X. Zhang, Phys. Rev. Lett., \textbf{99}, 153901.(2007).
\bibitem{sr23}A. Marini, A. V. Gorbach, and D. V. Skryabin, Opt. Lett., \textbf{35}, 3532.(2010).
\bibitem{sr24}A. Marini, D. V. Skryabin, and B. Malomed, Opt. Express, \textbf{19}, 6616.(2011).

\bibitem{sr25}F. Ye, D. Mihalache, B. Hu, and N. C. Panoiu, Phys. Rev. Lett., \textbf{104}, 106802.(2010).
\bibitem{ye1}X. Shi, X. Chen, B. A. Malomed, N. C. Panoiu, and F. Ye, Phys. Rev. B, \textbf{89}, 195428.(2014).
\bibitem{ye2}Y. Kou, F. Ye, and X. Chen, Phys. Rev. A, \textbf{84}, 033855.(2011).

\bibitem{egorov}O. A. Egorov, and F. Lederer, Opt. Lett., \textbf{38}, 1010.(2013).

\bibitem{noskov2013}R. E. Noskov, D. A. Smirnova, and Y. S. Kivshar, Opt. Lett., \textbf{38}, 2554.(2013).

\bibitem{prl}R. E. Noskov, P. A. Belov, and Y. S. Kivshar, Phys. Rev. Lett., \textbf{108}, 093901.(2012).
\bibitem{oe}R. E. Noskov, P. A. Belov, and Y. S. Kivshar, Opt. Express, \textbf{20}, 2733.(2012).
\bibitem{sr}R. E. Noskov, P. A. Belov, and Y. S. Kivshar, Sci. Rep., \textbf{2}, 873. (2012).


\bibitem{oe18}P. B. Johnson, and R. W. Christy, Phys. Rev. B, \textbf{6}, 4370.(1972).

\bibitem{oe20}V. P. Drachev, A. K. Buin, H. Nakotte, and V. M. Shalaev, Nano Lett., \textbf{4}, 1535.(2004).

\bibitem{clerc}M. G. Clerc, R. G. El\'{i}as, and R. G. Rojas, Phil. Trans. R. Soc. A, \textbf{369}, 412.(2011).

\bibitem{ishimori}Y. Ishimori, and T. Munakata, J. Phys. Soc. Jpn., \textbf{51}, 3367.(1982).

\bibitem{peyard}M. Peyard, and M. D. Kruskal, Physica D, \textbf{14}, 88.(1984).

\bibitem{karpan}V. M. Karpan, Y. Zolotaryuk, P. L. Christiansen, and A. V. Zolotaryuk, Phys. Rev. E, \textbf{66}, 066603.(2002).



\end{thebibliography}

\end{document}